\documentclass[
reprint,
superscriptaddress,
showpacs,preprintnumbers,
 amsmath,amssymb,
 aps,
 prc,
 floatfix,
 nofootinbib
]{revtex4-1}
\usepackage[T1]{fontenc}
\usepackage[applemac]{inputenc}
\usepackage[pdftex]{color}
\usepackage{babel}
\usepackage{amssymb}
\usepackage[pdftex]{graphicx}
\usepackage{tabularx}
\usepackage{footnote}
\usepackage{mathrsfs}
\usepackage[unicode=true,pdfusetitle,
 bookmarks=true,bookmarksnumbered=false,bookmarksopen=false,
 breaklinks=false,pdfborder={0 0 1},backref=false,colorlinks=true]
 {hyperref}
\hypersetup{
 urlcolor=blue, citecolor=blue}

\usepackage{dcolumn}
\newcolumntype{d}{D{.}{.}{1} }

\usepackage{lineno}%% for line number
\usepackage{xcolor}

\makeatletter
\@ifundefined{textcolor}{}
{%
 \definecolor{BLACK}{gray}{0}
 \definecolor{WHITE}{gray}{1}
 \definecolor{RED}{rgb}{1,0,0}
 \definecolor{GREEN}{rgb}{0,1,0}
 \definecolor{BLUE}{rgb}{0,0,1}
 \definecolor{CYAN}{cmyk}{1,0,0,0}
 \definecolor{MAGENTA}{cmyk}{0,1,0,0}
 \definecolor{YELLOW}{cmyk}{0,0,1,0}
}
\makeatother
\hypersetup{urlcolor=blue}

\usepackage{amsmath}

\newcommand{\isotope}[2]{$^{#1}$\text{#2}}

\begin{document}

%\preprint{APS/123-QED}

\title{Astrophysical significance of the isomer \isotope{119 \rm m}{Ag} demonstrated through direct mass measurement}

\author{F.~Rivero}
\email{frivero@nd.edu}
\affiliation{Department of Physics and Astronomy, University of Notre Dame, Notre Dame, IN 46556, USA}

\author{M.~Brodeur}
\affiliation{Department of Physics and Astronomy, University of Notre Dame, Notre Dame, IN 46556, USA}

\author{J.A.~Clark}
\affiliation{Physics Division, Argonne National Laboratory, Lemont, IL 60439, USA}

\author{B.~Liu}
\affiliation{Department of Physics and Astronomy, University of Notre Dame, Notre Dame, IN 46556, USA}
\affiliation{Physics Division, Argonne National Laboratory, Lemont, IL 60439, USA}

\author{G.W.~Misch}
\affiliation{Theoretical Division, Los Alamos National Laboratory, Los Alamos, NM, 87545, USA}

\author{M.R.~Mumpower}
\affiliation{Theoretical Division, Los Alamos National Laboratory, Los Alamos, NM, 87545, USA}
\affiliation{Center for Theoretical Astrophysics, Los Alamos National Laboratory, Los Alamos, NM, 87545, USA}

\author{W.S.~Porter}
\affiliation{Department of Physics and Astronomy, University of Notre Dame, Notre Dame, IN 46556, USA}

\author{D.~Ray}
\affiliation{Department of Physics and Astronomy, University of Manitoba, Winnipeg, MB R3T 2N2, Canada }
\affiliation{Physics Division, Argonne National Laboratory, Lemont, IL 60439, USA}

\author{G.~Savard}
\affiliation{Physics Division, Argonne National Laboratory, Lemont, IL 60439, USA}
\affiliation{Department of Physics, University of Chicago, Chicago, IL 60637, USA}

\author{T.M.~Sprouse}
\affiliation{Theoretical Division, Los Alamos National Laboratory, Los Alamos, NM, 87545, USA}

\author{A.A.~Valverde}
\affiliation{Department of Physics and Astronomy, University of Manitoba, Winnipeg, MB R3T 2N2, Canada }
\affiliation{Physics Division, Argonne National Laboratory, Lemont, IL 60439, USA}

\author{D.P.~Burdette}
\affiliation{Physics Division, Argonne National Laboratory, Lemont, IL 60439, USA}

\author{A.~Cannon}
\affiliation{Department of Physics and astronomy, University of Notre Dame, Notre Dame, IN 46556, USA}

\author{A.T.~Gallant}
\affiliation{Nuclear and Chemical Sciences Division, Lawrence Livermore National Laboratory, Livermore, CA 94550, USA}

\author{A.M.~Houff}
\affiliation{Department of Physics and Astronomy, University of Notre Dame, Notre Dame, IN 46556, USA}

\author{K.~Kolos}
\affiliation{Nuclear and Chemical Sciences Division, Lawrence Livermore National Laboratory, Livermore, CA 94550, USA}

\author{F.G.~Kondev}
\affiliation{Physics Division, Argonne National Laboratory, Lemont, IL 60439, USA}

\author{R.~Orford}
\affiliation{Nuclear Science Division, Lawrence Berkeley National Laboratory, Berkeley, California 94720, USA}

\author{C.~Quick}
\affiliation{Department of Physics and Astronomy, University of Notre Dame, Notre Dame, IN 46556, USA}

\author{K.S.~Sharma}
\affiliation{Department of Physics and Astronomy, University of Manitoba, Winnipeg, MB R3T 2N2, Canada }

\author{L.~Varriano}
\affiliation{Center for Experimental Nuclear Physics and Astrophysics, University of Washington, Seattle, WA 98195, USA}

\begin{abstract}

The abundance of elements heavier than iron produced via the astrophysical rapid-neutron capture process depends sensitively on the atomic mass of the involved nuclei as well as the behavior of a few special types of nuclear isomers called ``astromers''. High-precision mass measurements of \isotope{119}{Cd}, \isotope{119}{Ag} and their respective isomeric states have been performed with the Phase Imaging-Ion Cyclotron Resonance (PI-ICR) method with a precision of $\delta m/m \approx 10^{-8}$ using the Canadian Penning Trap (CPT). The ground state mass excess, as well as the excitation energy, agrees with recent Penning Trap measurements from JYFLTRAP. Network calculations using these new measurements revealed that, contrary to previous expectations, \isotope{119m}{Ag} behaves as an astromer which significantly affects the population of \isotope{119}{Ag}.

%The masses of neutron rich nuclei play a significnt role in the evolution of elements in the \textit{r} process and s-process. Precision values for these nuclear masses are necessary for accurate modeling in r- and s-process simulations. The method of Phase Imaging-Ion Cyclotron Resonance (PI-ICR) allows for measurement with a precision of $\delta m/m \approx 10^{-7}$. Here the Canadian Penning Trap (CPT) at Argonne National Lab was used to measure the mass of the ground state and isomer of $^{119}$Cd. 

\end{abstract}

\maketitle{}

\textit{Introduction - }Determining the exact astrophysical environment(s) responsible for the synthesis of about half of all isotopes heavier than iron remains a significant open question in physics \cite{Arnould2007} that requires efforts from observational astronomy \cite{Siegel2022}, galactic chemical evolution modeling \cite{Cote2019}, as well as experimental and theoretical nuclear physics \cite{Kajino2019}. This large number of isotopes is believed to be produced by the rapid neutron capture process (\textit{r} process) \cite{Arnould2007, Qian2003, Burbidge1957, Cowan1991}, which consists of a series of neutron captures and $\beta$-decays forming a path through the neutron-rich region of the chart of the nuclides. The observation of the neutron star merger GW170817, through its gravitational wave emission \cite{Abbott2017} and its electromagnetic spectra \cite{Chornock2017}, strongly supports the hypothesis of this astrophysical event as a location for the \textit{r} process \cite{ Kasliwal2017}. The confirmation of this astrophysical site increased the need for reliable data for nuclear physics observables to better understand how elements are formed and ejected during a merger event, as well as to determine if the \textit{r} process occurs in other environments \cite{Kajino2019}.

Detailed studies have revealed that, among the various nuclear physics observables of the nuclei involved, the calculated abundance produced by the \textit{r} process is most sensitive to atomic masses \cite{Mumpower2016}. This sensitivity has been confirmed by several recent mass measurements in various regions of the nuclear chart, resulting in substantial changes in the calculated abundances \cite{Izzo2021, Orford2022}. Recently, the role of nuclear isomers has also begun to be explored \cite{Misch2021, Misch2024}. 

These excited states can sometimes be very long-lived because of their angular momentum difference with the ground state, which suppresses electromagnetic transitions. However, in high-temperature astrophysical environments, such as during a neutron star merger, these isomeric states can couple to the ground state via intermediate states through processes such as thermal excitations, beta decay, neutron capture, and fission. As a result, the effective lifetime of an isotope can be altered by several orders of magnitude \cite{Aprahamian2005, Hoff2023}. Isomers which exhibit a strong impact on the abundances of nuclear species, denoted as ``astromers'' in \cite{Misch2020}, can have a large influence on the \textit{r} process path and abundance pattern.

 Silver is known to have several neutron-rich isotopes that present one or even multiple isomeric states \cite{Kondev2021}. One such isotope of interest is \isotope{119}{Ag}, whose isomeric state has an unknown excitation energy \cite{Kondev2021}. Similarly to the mass excess of the ground state, precise and accurate knowledge of excitation energies is an essential ingredient for astrophysical \textit{r} process network calculations. Consequently, the atomic masses of both states of \isotope{119}{Ag} as well as its daughter nucleus, \isotope{119}{Cd}, which are both part of the $\beta$-decay chain that populates the $Z=50$ proton shell closure at \isotope{119}{Sn}, were measured using the Canadian Penning Trap mass spectrometer (CPT) at the CAlifornium Rare Isotope Breeder Upgrade (CARIBU) facility \cite{Schelt2013} at Argonne National Laboratory (ANL).

\textit{Experimental Facility - }Studies of exotic nuclei with the CPT begin with a radioactive ion beam (RIB) produced by CARIBU \cite{Savard2008}, which consists of a spontaneously-fissioning $\approx$1 Ci \isotope{252}{Cf} source. The source is encased in a large volume gas catcher containing ultra-pure helium, with a pressure on the order of 35 torr. Fission fragments must first pass through a thin gold foil that removes most of their kinetic energy. Upon entering the gas region, they are thermalized via elastic collisions with the helium gas. These low-energy ions are extracted with a combination of static and dynamic electric fields, as well as outward gas flow, through a small exit nozzle, into a differentially pumped radio-frequency quadrupole (RFQ) ion guide. From there, the ions are sent through an isobar separator with a mass resolution of $m/\Delta m \approx 14,000$ \cite{Orford2018}, which selects the appropriate $A/q$ ratio, 119/1+ in this case. The ions then enter an RFQ buncher that cools and bunches the beam, changing it from a continuous stream of ions to a set of discrete bunches with a well-defined energy.

From here the ions enter a multi-reflection time-of-flight (MR-TOF) mass spectrometer with mass resolution on the order of $m/\Delta m \approx 10^5$ \cite{Hirsh2016} where some of the previously unresolved isobars can be separated by time-of-flight. Finally, a Bradbury-Nielsen gate (BNG) \cite{Bradbury1936} is used to select the species of interest. At this point, the beam is almost entirely composed of \isotope{119}{Ag} and \isotope{119}{Cd}, with minimal contaminants, which are mainly molecular.

After the BNG, the ions enter the CPT system, which consists of a voltage elevator to decrease the beam energy, a linear Paul trap to prepare the ions for injection into the CPT, and the measurement Penning trap which is housed in a 5.7 T superconducting solenoid magnet \cite{Gamage2022}. In the trap, the ions are subjected to several excitation pulses that are applied during the measurement of their cyclotron frequency using the phase-imaging ion-cyclotron resonance (PI-ICR) technique \cite{Eliseev2013}.

\begin{figure}
    \centering
    \includegraphics[width=\linewidth]{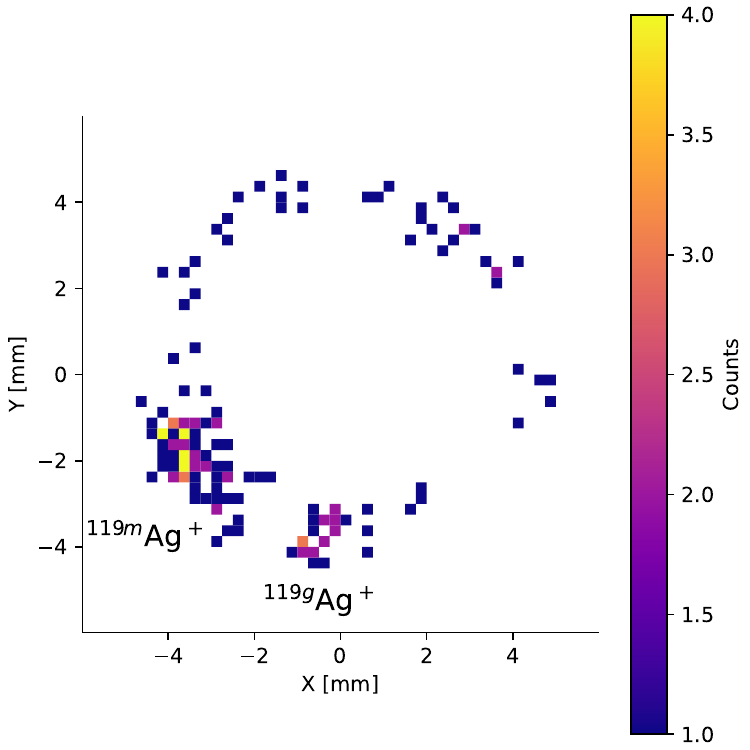}
    \caption{Typical 2D histogram of ion hits on the PS-MCP after a final measurement phase. This image is from a measurement of the ground state of \isotope{119}{Ag} with $t_{acc}$=710.324 ms. During this particular run, the PS-MCP would typically record a single ion per trap ejection, and the spots shown in this plot are those where fewer than 3 ions hit the PS-MCP in a single trap ejection. The ground state and isomer states of Ag are clearly separated.}
    \label{fig:mcp}
\end{figure}

\textit{Measurement Techniques - }The PI-ICR excitation scheme used at the CPT \cite{Orford2020} consists of two modes: reference phase measurement, which defines a starting point for phase accumulation, and final phase measurement, which allows for a larger phase accumulation. When an ion bunch is captured inside the trap, \textit{hot} ions are first removed by adiabatically raising the potential of the ring electrode. The remaining ions are prepared in a reduced cyclotron motion orbit using a dipole excitation. For a reference run, the reduced cyclotron motion is then immediately converted to magnetron motion via a broadband quadrupole signal at the cyclotron frequency of the ion of interest. After a brief time, the bunch is ejected from the trap and its position is recorded on a position-sensitive micro-channel plate (PS-MCP) detector. This position is used to mark the starting point, or reference spot, of phase accumulation. For final phase measurement runs, there is a period of time between the dipole and quadrupole excitations, referred to as $t_{acc}$, during which the ions in the trap accumulate some amount of phase according to their mass. After the quadrupole excitation, the ion remains in the trap for some amount of time such that the total amount of time spent in the trap for measurement runs and reference runs is equal. With an appropriately selected accumulation time, the different species present in the trap (and their isomeric states) can be resolved, as shown in Fig. \ref{fig:mcp}. A Gaussian mixture model code was used to group the spots into clusters and determine their centroid, from which a phase value could be obtained \cite{Weber2022}. The cyclotron frequency of the ion of interest is then calculated from the phase difference between the two excitation schemes, the number of revolutions, and the accumulation time.

Despite the robust mass resolving power of the system as a whole, small amounts of contaminants may still make it into the Penning trap. By testing multiple accumulation times, the cyclotron frequency of these contaminants can be determined. In doing this, nearby isobaric contaminants can be readily identified. However, the use of a large volume gas cell to stop the decay products of the \isotope{252}{Cf} source generates molecular contamination. The veil of ions present in the top right portion of the excitation ring in Fig. \ref{fig:mcp} could be contaminants. Unfortunately it is in an amount too small to be unambiguously identified.

Finally, from the simple relationship for the cyclotron frequency
\begin{equation}
    \nu_c = \frac{1}{2\pi} \frac{qB}{m_{ion}}
\end{equation}
one can extract the mass, $m_{ion}$, of the ion of interest, with a known charge, $q$, and magnetic field, $B$.  

To account for possible variations in the magnetic field, it is calibrated by measuring the cyclotron frequency of an ion, $\nu_{c,cal}$, with a very precisely known mass, shortly after the ion of interest. From the frequency ratio $R = \nu_{c,cal}/\nu_c$, the atomic mass $m$ of the species of interest is found using 
\begin{equation}
    m = R \frac{q}{q_{cal}} (m_{cal} - q_{cal}m_e) + q m_e
\end{equation}
where $cal$ denote the calibrant species, $q$'s are charge states, and $m_e$ is the electron mass. For all measurements in this work, \isotope{133}{Cs$^+$} was used as the calibrant. The electron binding energies, on the order of eVs, are neglected.

\begin{figure}
    \centering
    \includegraphics[width=\linewidth]{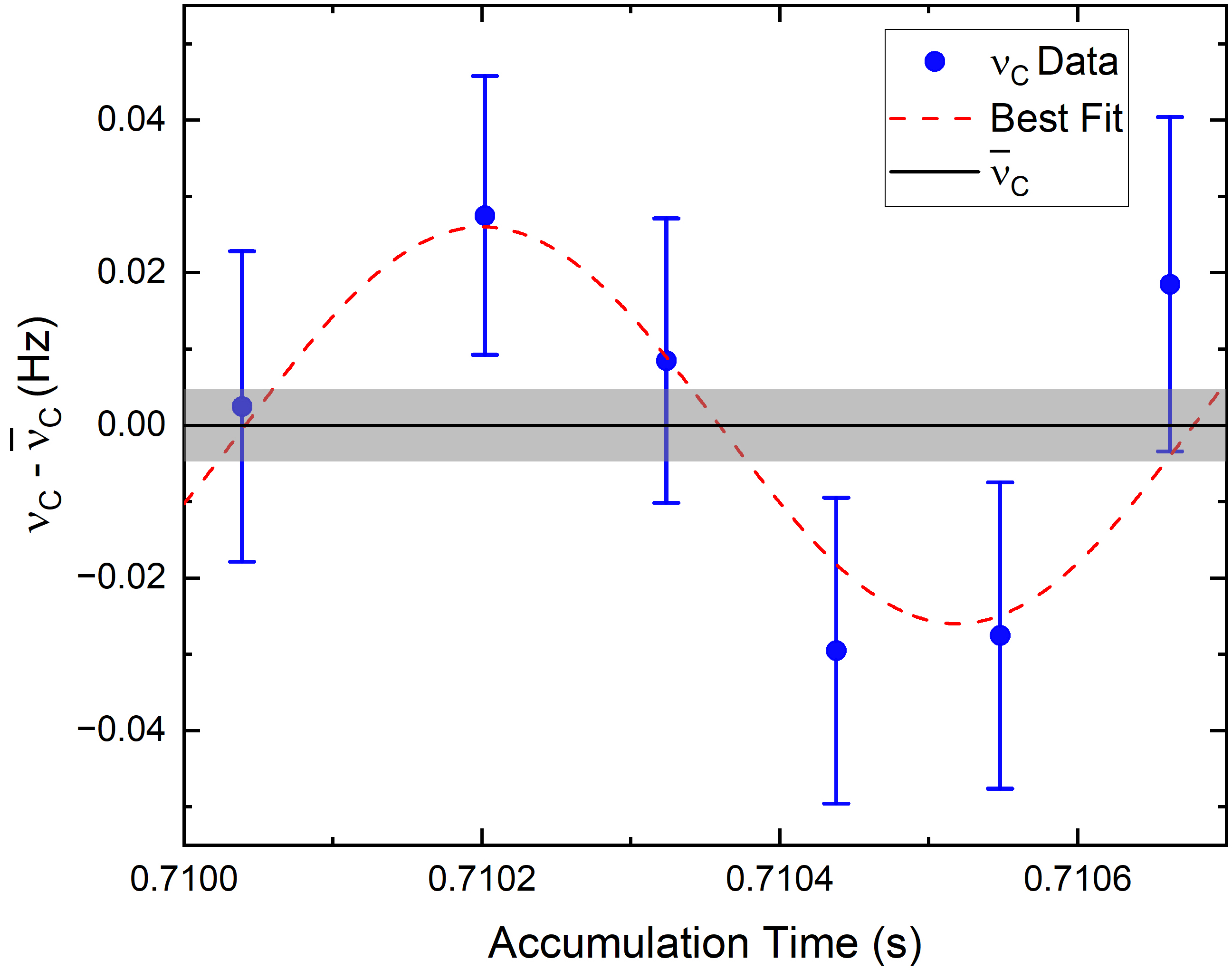}
    \caption{Data (in blue) and sinusoidal fit (in red) of the \isotope{119}{Ag$^+$} ground state measurement, using the model in \cite{Orford2020}. The dark gray line represents the true cyclotron frequency, and the lighter gray band is the  statistical uncertainty.}
    \label{fig:gsSine}
\end{figure}

\textit{Systematics - }There are several types of systematic effects present that can affect the determination of the cyclotron frequency. The first and largest is a residual magnetron motion present at the start of the dipole excitation \cite{Orford2020}. This effect results in a sinusoidal variation of the measured cyclotron frequency as a function of the accumulation time with a period equal to the inverse of the magnetron frequency $\nu_-$. As such, the true cyclotron frequency is determined using the process described in \cite{Orford2020}, where several measurements of $\nu_c$ are performed within a period of $\nu_-$. These measurements are fit using a sinusoidal curve whose mean value is the true cyclotron frequency. Figure \ref{fig:gsSine} shows this sine curve for \isotope{119g}{Ag}. The second systematic is the small, but nonzero, phase that different isotopes will have within a reference spot. To correct for this, an iterative correction method has been used \cite{Orford2020} resulting in changes in the cyclotron frequency of less than 10 ppb. The measured cyclotron frequency was also observed to depend on the recorded position of the ions on the PS-MCP. This effect has been corrected following the method described in \cite{Liu2024} and is also on the order of 10 ppb.

Further systematic effects include distortion of the trapping potential, misalignment of the trap and magnetic field axis \cite{Brown1982}, ion-ion interaction \cite{Bollen1992} and relativistic effects \cite{Brodeur2009}. The relativistic effect is negligible at the lower frequencies of the heavy species involved in this work. Shifts due to ion-ion interaction are important when different species are present in the trap; however, this effect was minimized in this analysis to less than 1 ppb by limiting the accepted signals to those where less than 4 ions hit the PS-MCP in a single shot \cite{Ray2024}. Systematic effects like magnetic field inhomogeneities, misalignment of the trap and magnetic field axes, and non-harmonic terms in the trap potential all result in a shift that scales linearly with the frequency ratio of two species of sufficiently different mass, and for that reason takes the name of ``mass-dependent shift''. That effect was investigated by calculating the frequency ratio of pairs of stable isotopes of different mass separation and was found to be $\Delta R / R = 4.1(17)\times 10^{-10} \cdot  ({{m}/{q}-{m_{cal}}/{q_{cal}}})$. The size of this quantity was conservatively added as a systematic uncertainty rather than a correction. 

Potential systematics arising from temporal instabilities in magnetic or electric fields, non-circular projection on the MCP, and various other effects have been studied in \cite{Ray2024}, and have been determined to have an effect smaller than 4 ppb. This value has also been added in quadrature to the uncertainty.

%which makes use of the relative abundance of each species within the trap, is used. It assumes, first, the correct measured cyclotron frequency, then corrects based on relative population fraction. This process is repeated until the change in frequency is less than the statistical uncertainty. This correction is often less than 5 kev/$c^2$ \cite{Orford2020}. 

%The ions are allowed to oscillate at their mass-dependent cyclotron frequency for a fixed amount of time, and complete a known number of loops around the trap region.

\begin{figure}
    \centering
    \includegraphics[width=\linewidth]{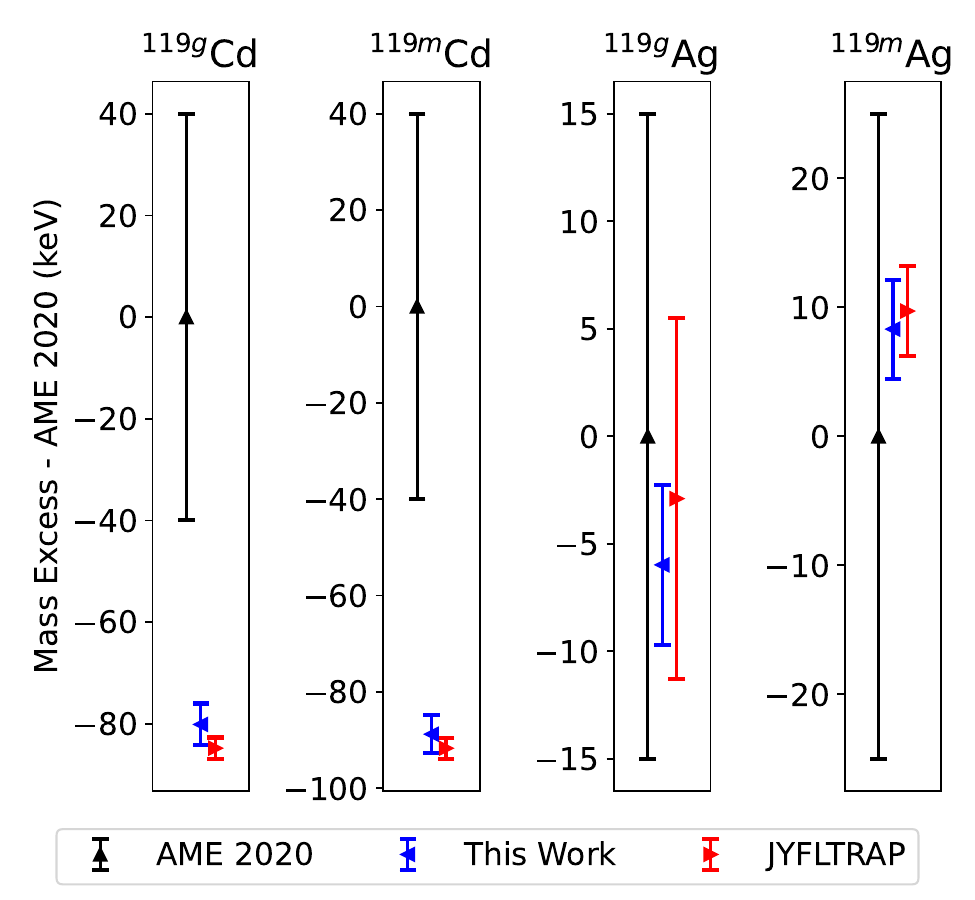}
    \caption{Mass excesses of the ground state and isomeric states of \isotope{119}{Cd} and \isotope{119}{Ag} as compared to the AME2020 value. JYFLTRAP measurements for Cd and Ag are from \cite{Jaries2023} and \cite{deGroote2024}, respectively. }
    \label{fig:ME}
\end{figure}

\begin{figure}
    \centering
    \includegraphics[width=\linewidth]{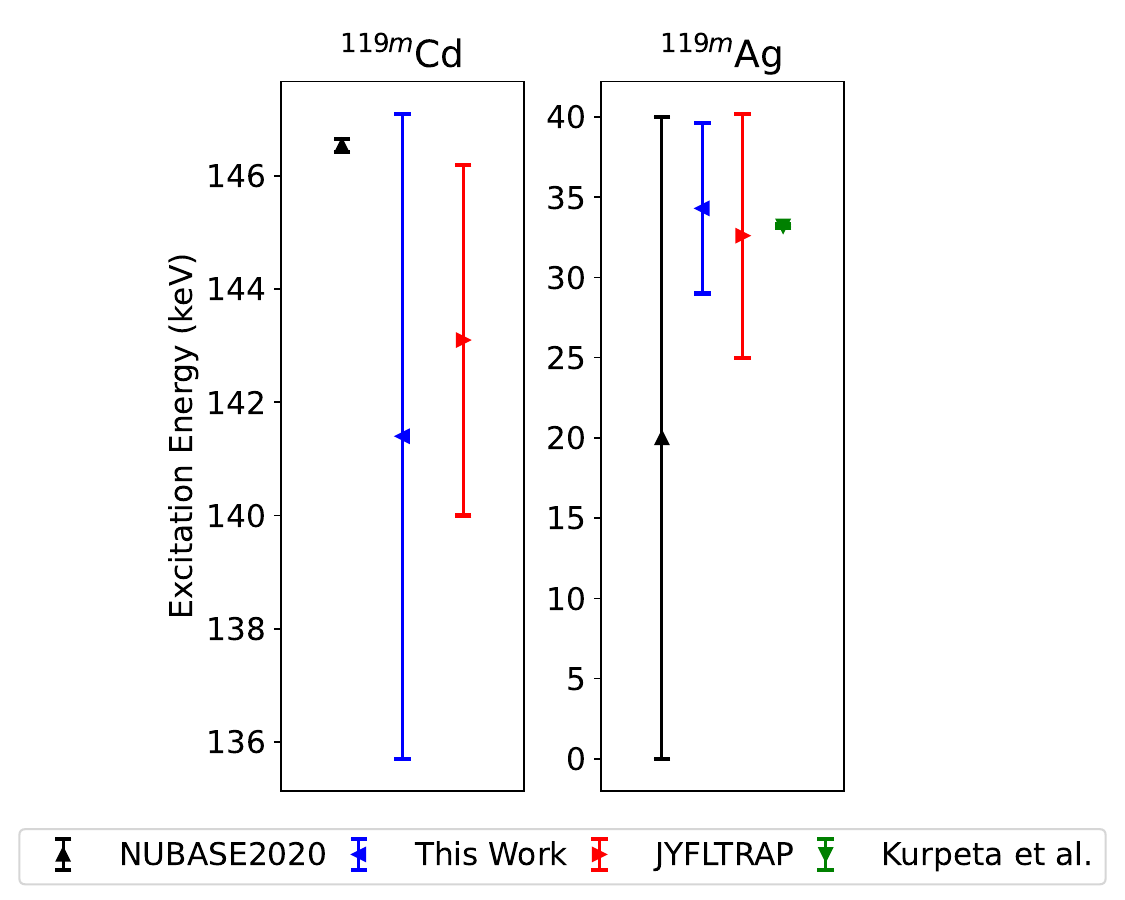}
    \caption{Isomer excitation energies for \isotope{119}{Cd} and \isotope{119}{Ag}, relative to their respective ground states, as determined by this work, NUBASE 2020, and recent publications.}
    \label{fig:ExEn}
\end{figure}

\textit{Results - }Table \ref{tab:ME} lists a summary of the results from this work as well as values from the 2020 atomic mass evaluation (AME2020) \cite{AME2020}, the recent mass measurements of \isotope{119}{Cd} and \isotope{119}{Ag} performed by the JYFLTRAP Penning trap \cite{Jaries2023, deGroote2024}, and the calculation of the excitation energy of \isotope{119}{Ag} by \cite{Kurpeta2022}, obtained with $\gamma\gamma$- and $\beta\gamma$-spectroscopy, complemented by a prompt-$\gamma$ study of levels in \isotope{119}{Ag}. These results are also presented graphically in Fig.~\ref{fig:ME}. Previously, the JYFLTRAP group observed a dramatic shift of $\approx$80 and $\approx$90 keV, for the Cd ground state and isomer respectively, with respect to the value of AME2020, which was obtained from a $\beta$-endpoint measurement \cite{Aleklett1982}. This shift is now confirmed by this work. Meanwhile, the excitation energy of \isotope{119m}{Cd}, as determined by this work, the AME2020, and JYFLTRAP, maintains a good agreement, as shown in Fig. \ref{fig:ExEn}. 

The \isotope{119}{Ag} measurements, on the other hand, agree well with the AME2020 values and provide a significant improvement to their precision, being 1.8 and 7.1 times more precise than the value in the AME2020 for both the ground and isomer states, respectively, making this the most precise resolution of the isomer state obtained from direct mass measurements. 

\begin{table*}[!ht]
\caption{\label{tab:ME}
Cyclotron frequency ratios with respect to \isotope{133}{Cs$^+$} ($R$), mass excesses ME, and isomer excitation energies $E_x$ for \isotope{119}{Cd$^+$} and \isotope{119}{Ag$^+$} as obtained in this work, the AME2020, and recent publications. }
    \centering
    \begin{tabular}{ccccccccc}
        \hline
        ~  Isotope  & $R$ & ME (keV) & $\rm ME_{lit.}$ (keV) & ME$_{\text{JYF}}$ (keV) & E$_{x}$(keV) & E$_{x, lit}$(keV) & E$_{x, \text{JYF}}$(keV) & E$_{x, \text{Kurpeta}}$(keV)\\ 
        \hline
        \hline
         \isotope{119g}{Cd} & 0.894 693 921(33) & -84060.1(41) & -83980(40)\footnotemark[1] & -84064.8(21)\footnotemark[3] & - & - & - & -\\
         \isotope{119m}{Cd} & 0.894 695 063(32) & -83918.8(39) & -83830(40)\footnotemark[2] & -83921.7(22)\footnotemark[3] & 141.4(57) & 146.54(11)\footnotemark[2] & 143.1(31)\footnotemark[3]& -\\ 
         \isotope{119g}{Ag} & 0.894 737 606(30) & -78652.0(37) & -78646(15)\footnotemark[1] & -78648.9(84)\footnotemark[4] & - & - & - & -\\ 
         \isotope{119m}{Ag} & 0.894 737 883(31) & -78617.7(38) & -78626\text{\#}(25\text{\#})\footnotemark[2] & -78616.3(35)\footnotemark[4] & 34.3(53) & 20\#(20\#)\footnotemark[2] & 32.6(76)\footnotemark[4] & 33.4(1)\footnotemark[5]\\ 
        \hline
    \end{tabular}
    \footnotetext[1]{From the 2020 Atomic Mass Evaluation \cite{AME2020}}
    \footnotetext[2]{From 2020 NUBASE \cite{Kondev2021}}
    \footnotetext[3]{From Jaries et al. \cite{Jaries2023}}
    \footnotetext[4]{From de Groote et al. \cite{deGroote2024}}
    \footnotetext[5]{From Kurpeta et al. \cite{Kurpeta2022}}
\end{table*}

\textit{Discussion - }The importance of \isotope{119 \rm m}{Cd} and \isotope{119 \rm m}{Ag} as possible astromers has been studied by performing network calculations in the presence and absence of these isotopes using a weighted average of the mass excesses presented in this work and those from the JYFLTRAP group \cite{Jaries2023, deGroote2024}, and the computational machinery of \cite{Misch2021} and \cite{Sprouse2022}.

\begin{figure}
    \centering
    \includegraphics[width=1.03\linewidth]{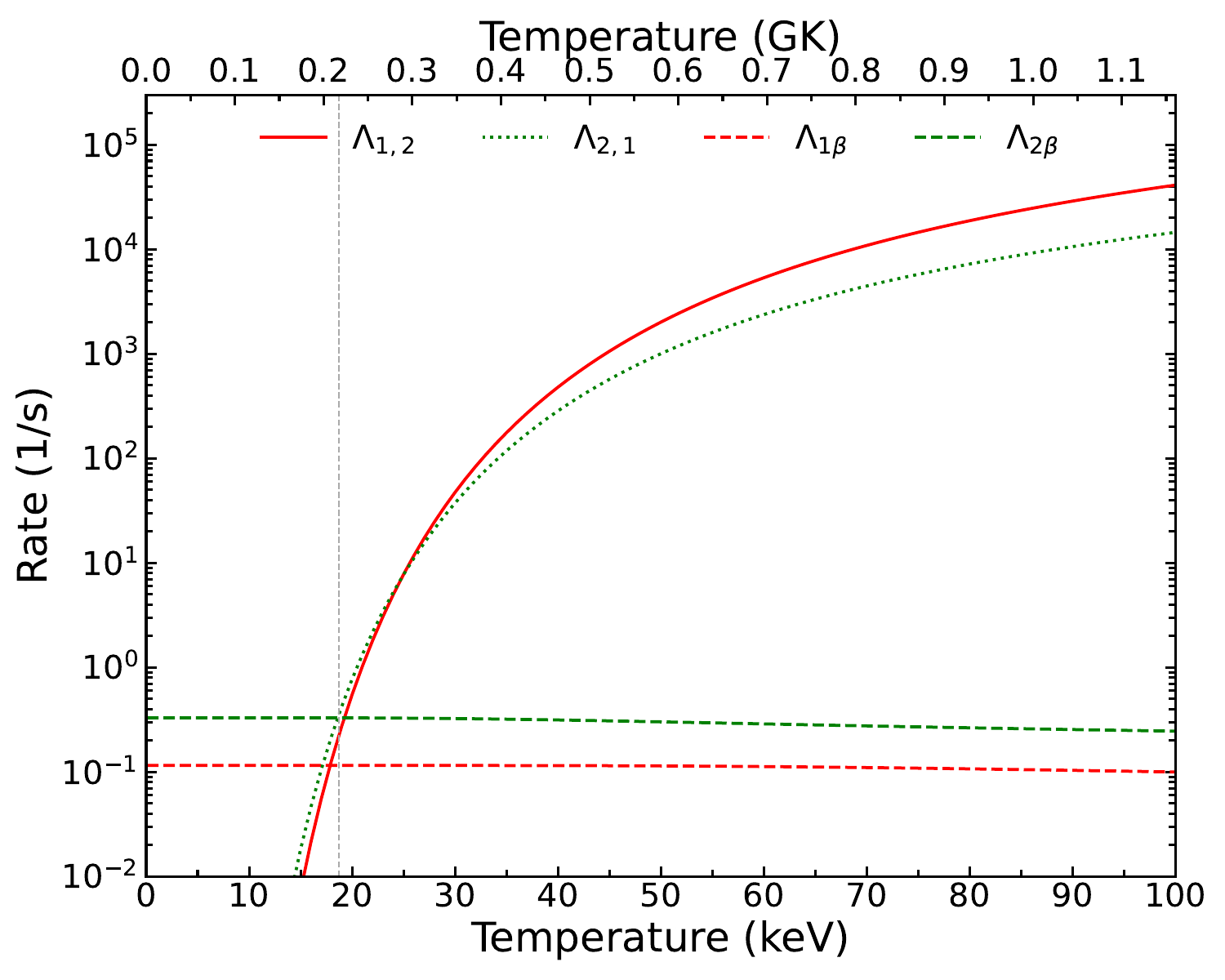}
    \caption{Reaction rates used in the presented calculations as function of thermalization temperature. Shown are the rate of thermal excitation of the ground state $\Lambda_{1,2}$ (solid red), the effective de-excitation rate from the isomeric state $\Lambda_{2,1}$ (dotted green), as well as the respective $\beta$-decay rates of the ground state ($\Lambda_{1 \beta}$, dashed red), and isomer state ($\Lambda_{2 \beta}$, dashed green). The vertical dashed line at 18.7 keV indicates the temperature below which thermal equilibrium between the two states breaks.}
    \label{fig:reactionRates}
\end{figure}

Using the results of the \textit{r} process abundance pattern obtained from a simulation of the typical conditions present in the accretion disc of a binary neutron star merger \cite{Sprouse2024}, the reaction rates and abundances of the measured isotopes were calculated as per the process described in \cite{Misch2021,Sprouse2022, Misch2020}. The new excitation energy measurement of \isotope{119 \rm m}{Cd} did not change the isomeric impact of this nucleus, since the energy of the isomeric state remained the same with respect to the ground state. The measurements of \isotope{119}{Ag} from this work and from \cite{deGroote2024, Kurpeta2022} proved to be more impactful since they provided the improved resolution of the ground and isomeric states, along with the respective excited states built on these two levels. We now provide the first estimate of the beta-decay thermalization temperature for this nucleus, roughly 18.7 keV, as shown in Fig.\ref{fig:reactionRates}. The possible discovery of additional levels in this nucleus may alter this value. Figure~\ref{fig:reactionRates} shows the rates used in the decay network in terms of thermalization temperature. At temperatures below 18.7 keV, the rate of thermal stimulation from the isomer to the ground state falls below the beta-decay rate of the isomeric state, breaking the thermal equilibrium with the ground state. Consequently, \isotope{119}{Ag} can no longer be treated as a single species in network calculations, and its isomeric state is considered to be an astromer. 

\begin{figure}
    \centering
    \includegraphics[width=1.03\linewidth]{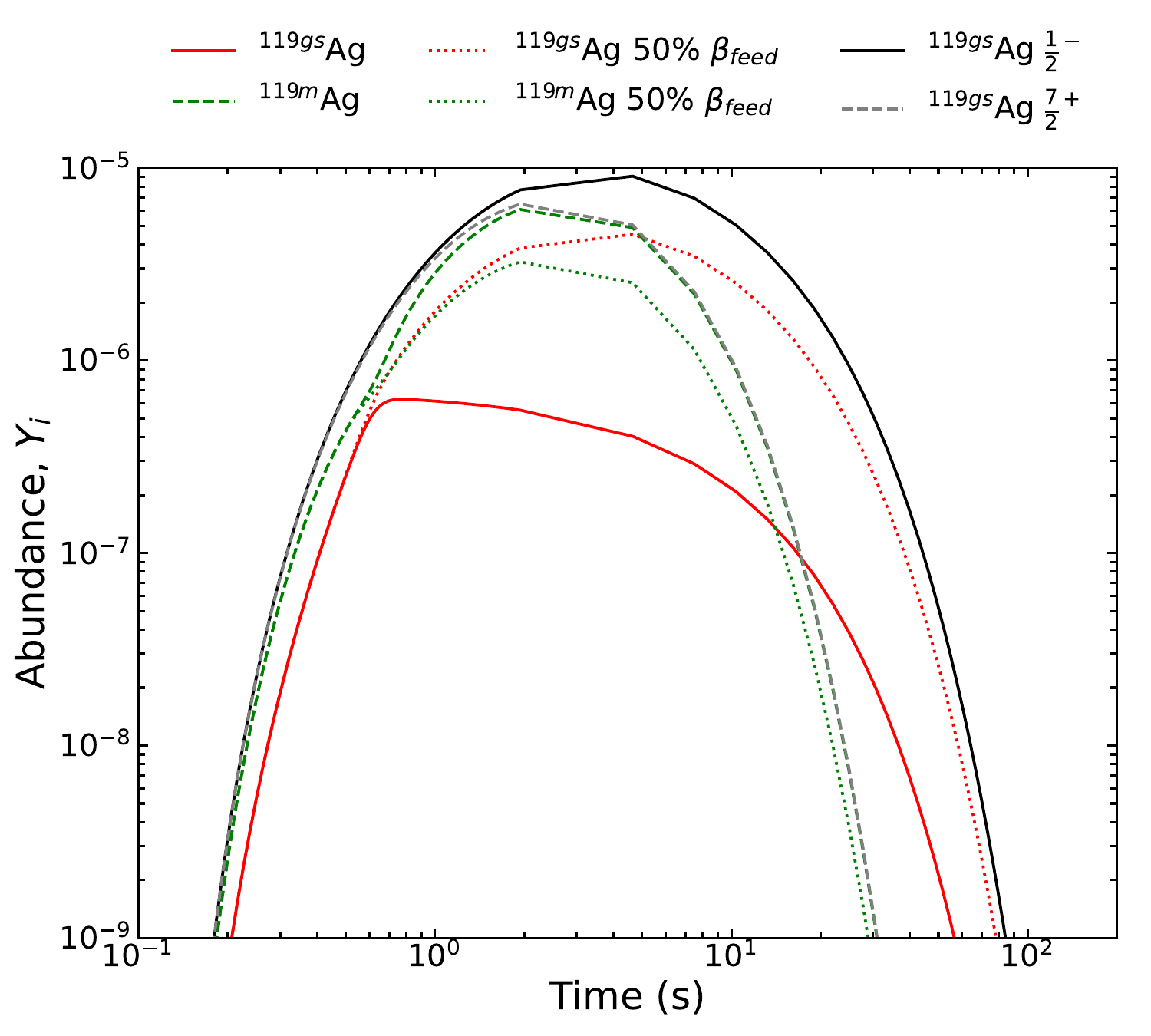}
    \caption{Abundance of the two states of $^{119}$Ag as a funtion of simulation time. The solid red and dashed green curves assume no feeding from $^{119}$Pd to the ground state. The two dotted curves represent the abundances of the ground and isomer states assuming a 50\% feeding rate to each of those states from the beta-decay of \isotope{119}{Pd}. Finally, the solid black and dashed gray curves adopt the treatment of \isotope{119}{Ag} as a single state with $J^\pi= 1/2^-$ or $7/2^+$, respectively.}
    \label{fig:abundances}
\end{figure}

Figure~\ref{fig:abundances} shows the behavior of the abundance of the states of this nucleus as a function of time in the \textit{r} process. Due to the higher spin of the isomer ($J^\pi=7/2^+$), this state is favored by thermal equilibrium at early times. At around 0.6 seconds, the temperature drops below 18.7~keV, the thermal equilibrium breaks down, and the ground state is no longer populated by transitions from the isomer, marking the point at which \isotope{119m}{Ag} becomes an astromer. From this point, the population of the ground state follows either the solid red line, which assumes there is no feeding from the beta-decay of \isotope{119}{Pd}, or the dotted red line, which assumes 50\% feeding to the ground state. 

Currently, there is limited information available about \isotope{119}{Pd}. The only known levels are the ground state and an isomeric state, first proposed by \cite{Kurpeta2022}, and later measured directly with a mass-measurement Penning trap by \cite{Jaries2024}. The earlier study proposes a ground state of either $J^\pi=1/2^+$ or $3/2^+$, and a $11/2^-$ isomer. The later study then refined the isomeric excitation energy to be 199.1(30) keV. Due to the limited knowledge, about \isotope{119}{Pd}, it is difficult to make reliable predictions about the feeding rate of beta decay to the ground state vs the isomer of silver. 

The study done by \cite{Kurpeta2022} also reports a two-band structure in \isotope{119}{Ag}, corresponding to levels that mainly $\gamma$-decay to the ground state or the isomer. There are, however, three levels that can decay to both bands, allowing for the relative positioning of the energy levels. Both of these bands contain a number of excited states with low to medium spin ranging from $1/2$ to $13/2$, which prevents a precise determination of the feeding rate to each of the states in silver. For this reason, the network calculations also implemented a feeding rate of 0\% to the ground state. 

In the case of a 50\% feeding to both the ground and isomer bands, the population of the ground state begins to dominate almost immediately after thermal equilibrium breaks. The abundances stay relatively close until the population of \isotope{119}{Pd} is exhausted, at about $t=1 s$, and the two curves begin to diverge more dramatically. In the limiting case of 0\% feeding to the ground band, the simulation indicates that the isomer dominates the population of \isotope{119}{Ag} by about an order of magnitude, until about 20 s, when the shorter half-life of this state brings its abundance below that of the ground state.

This simulation considered two other limiting cases, which are also shown in Fig.~\ref{fig:abundances}. These are the treatment of \isotope{119}{Ag} as a single state, assuming uncertainty in the spin assignment, represented by the solid black and dashed gray curves, respectively. The gray curve may be difficult to see, given its overlap with the solid black and dashed green curves. The comparison of these two curves to the multi-state considerations indicates the effect of the isomer on the population of \isotope{119}{Ag}. This effect is more pronounced at longer times, where the population of the silver ground state is exhausted anywhere between $5-25 s$ earlier than would otherwise be the case.

\textit{Conclusion - }The masses of \isotope{119}{Cd} and \isotope{119}{Ag} as well as their isomeric states have been measured with the Canadian Penning Trap at the CARIBU facility of Argonne National Laboratory. The \isotope{119}{Cd} ground state mass excess, -84060.1(41) keV, and the excitation energy, 141.4(57) keV, are in good agreement with a recent measurement from JYFLTRAP. The \isotope{119}{Ag} ground state mass excess is in agreement with the AME2020 value based on a measurement of the ISOLTRAP group while being a factor of 5.3 times more precise. The \isotope{119}{Ag} isomeric state is clearly resolved and measured as 34.3(53) keV. Both states are also in good agreement with recent JYFLTRAP measurements. Using these new measurements, an astrophysical network calculation was performed which unambiguously established the astromeric nature of \isotope{119 \rm m}{Ag}. It is found that around 0.6 s after the conditions studied in this work, when the temperature drops below 18.7~keV, \isotope{119 \rm m}{Ag} becomes an astromer and must be treated separately from the ground state. This remains true until approximately 20~s, when the ground state abundance overtakes the isomeric state due to their differing half-lives. 

To better refine our understanding of the impact of this isomer and our understanding of the production of \isotope{119}{Ag} in astrophysical environments, spectroscopy of the states above the isomer in \isotope{119}{Ag}, as well as measurements of the feeding rate from \isotope{119}{Pd} to both the ground and isomeric states of \isotope{119}{Ag}, are necessary.

%Finally, a measurement of the unknown feeding of \isotope{119 \rm m}{Ag} from the beta-decay of \isotope{119}{Pd} would be desirable in the future to better constrain our understanding of the production of \isotope{119}{Ag} in astrophysical environments. 

\textit{Acknowledgments - }This work was supported in part by the National Science Foundation (NSF) Grant No.~PHY-2310059, U.S. Department of Energy by Lawrence Livermore National Laboratory under Contract DE-AC52-07NA27344, the US Department of Energy, and supported from the Natural Sciences and Engineering Research Council of Canada: Grant SAPPJ-2018-00028, as well as by the U.S. Department of Energy, Office of Nuclear Physics, under Contract No. DE-AC02-06CH11357 (ANL). L.V. was supported by the National Science Foundation Graduate Research Fellowship under Grant No. DGE-1746045. This research used resources of ANL's ATLAS facility, which is a DOE Office of Science User Facility. M.R.M. acknowledges support from the Laboratory Directed Research and Development program of LANL project number 20230052ER. G.W.M., M.R.M., and T.M.S. were supported by the US Department of Energy through LANL. LANL is operated by Triad National Security, LLC, for the National Nuclear Security Administration of U.S. Department of Energy (Contract No. 89233218CNA000001).

\bibliographystyle{apsrev4-2}
\bibliography{library}

\end{document}